\documentclass[12pt]{article}
\usepackage{amsmath,amssymb,amsfonts}

\newcommand{\ZZ}{\mathbb{Z}}
\newcommand{\EE}{\bold E}
\newcommand{\Symm}{\operatorname{Symm}}
\newcommand{\wh}{\widehat}
\newcommand{\wt}{\widetilde}

\begin{document}

\title{Thermo- and gas-dynamical processes in NPPs after
accidents} 
\author{V.~P.~Maslov}
\date{}

\maketitle

\begin{abstract}
 In the theory of superfluidity and superconductivity, 
a jump of the free energy was discovered theoretically
and was naturally called 
a {\it zeroth-order phase transition}.
 We present an example of an exactly solvable problem 
in which such a phase transition occurs.
\end{abstract}

\section{Introduction}

The Nuclear Power Plants (NPPs) cannot be stopped comletely 
after their resources have been exhausted.
And this fact of ``long-living'' means 
that the well-known decay microprocesses develop 
in the depths of the NPPs 
after their service time is over
or after an accident occurs.
We must study these microprocesses 
from the standpoint of macrophysics, 
i.e., we must use the averaged thermodynamical approach. 

At present, we have survived several accidents in NPPs
the greatest of which is the accident in the 4th block 
of the Chernobyl NPP. 
The resources of the Leningrad NPP will soon be over 
(and this NPP has already issued a radioactive release).
What shall we do further with this NPP? 
It was decided to continue its operation time.

The service times of the French NPPs
will also come to the end in the near future.
How will they behave after the reactor is shut down?
It is impossible to perform such experiments 
in laboratories. 

This is a completely new phenomenon, 
and we face a problem which has never been posed before:
to understand what the ``black box''
(an NPP very slowly dying after an accident 
or after the end of its service time) 
contains.

In the spring of 1986, 
I was invited to collect a group of mathematicians 
(and to be the head of this group)
whose task was to eliminate the effects 
of the accident in the 4th block of the Chernobyl NPP.
The enormous responsibility, very short terms, 
unexpectedness of some new phenomena 
observed for the first time 
demanded a huge strain of our forces 
and completely turned over my understanding of mathematics 
as a science.

In 1987 we published our report in the form of book~\cite{01},
but we did not find answers to the main questions.

It was clear that the modern mathematics 
must be revised in order to generalize and intensify 
the thermodynamics, the theory of probabilities, and 
the mathematical physics, i.e.,
the branches of science required to solve the problems 
arising in NPPs after accidents.

Since we cannot get inside a broken reactor 
and perform measurements there, 
we obtain only average results of measurements,
i.e., thermodynamical results. 
Inside the broken reactor, there may be no radiation at some
places, and dangerous spots are spread at random.
To explain several factors, we have to revise and generalize the
thermodynamical science with regard to unexpected physical phenomena.
Since the processes of nuclear decay continue in an NPP after
an accident or after its service time is over, we must consider
these processes from the thermodynamical standpoint assuming
that the ``service time'' of the NPP is infinitely long.

In general, it is already somewhat to late to revise 
the basic notions of the classical thermodynamics, 
because, after the superfluidity was discovered, 
the idea that all the molecules stop their motion 
at the temperature zero was refuted by the experiments.
Therefore, it is necessary to study the superfluidity 
from the physical standpoint and from the standpoint of 
physical-and-mathematical generalizations. 
It is necessary to generalize the remarkable results 
obtained by Landau's and Bogolyubov's schools so that 
it were possible to use them in the seemingly opposite
situations arising after accidents in NPPs.

The works of physicists of genius intuition 
must later be generalized from the mathematical-and-physical and,
perhaps, philosophical standpoint by theorists 
of such a high level as 
von Neumann, Wigner, and others. 

In his letters, A.~N.~Kolmogorov cites P.~L.~Kapitsa's opinion 
about superfluidity. 
Kolmogorov writes: 
``P.~L. (Kapitsa~-- Victor Maslov) reported the following:
already before the war, he (Kapitsa~-- Victor Maslov) 
performed a series of experiments for verifying 
Landau's theory. The results of these experiments 
(which had not been published) absolutely contradicted 
all Landau's constructions. \dots 
Obviously, the publications were delayed on purpose, 
so that Landau could not correct his theory 
(because, as is known, `he could prepare any theory
justifying any given data'!)''~\cite{02}, p.~594.

In fact, to give an instantaneous response 
to any new experimental data is a talent 
and a step towards the truth. 
And to relate the concept of superfluidity,
the classical thermodynamics, 
the theory of averaging 
(the theory of probabilities), 
and the data of accidents as a whole, 
this is already a problem for a mathematical physicist. 

I had a powerful stimulus to establish this relationship, 
because the problems that I could not solve at the time
of the Chernobyl accident have been bothering me
for the last 20 years, 
and the emotional experience I went through 
when signing the permission to close the sarcophagus
without being fully confident 
have been torturing me even in my sleep.
Each of the regular releases in the NPP 
returned me to these problems.

To solve this problem, we must combine 
the statistical physics and the science of its averaging in physics
(the thermodynamics),
as well as 
the mathematical statistics and the science of its averaging 
(the theory of probabilities),
as a single science.

But the modern theory of probabilities in not sufficient
to deal with the processes in NPPs. 
If the probability of an accident is negligibly small, 
but the effects of the accident are grandiose and disastrous,
then even an ``infinitely'' small probability cannot be neglected.

The second point is the following. 
Let us consider a mathematical example 
about the heads and tails play. 
The probability of a head or a tail at a given moment 
is $1/2$ independently of the prehistory of heads and tails 
till the present moment. But, as the famous English physicist 
R.~Peierls wrote about the roulette play 
(black and red instead of heads and tails),
it is doubtful that ``there would be at least one player who,
once taking the risk of black,
will continue to stake constantly the same number \dots 
the probability of the appearance of such a long ``black''
series is negligibly small so that it must be ignored completely''
\cite{03}, p.~126. 
How does this agree with the fact that the probability 
is independent of the prehistory?  

These considerations are very close to the notion 
of ``general position'' in topology. 
So, for example, the ``general position'' of a curve 
on the plane with respect to the projection 
on the $x$-axis is 
the position of the curve such that the ``cardinality'' 
of the set of nearby curves with the same properties 
is much greater than the ``cardinality'' of the set 
of curves with different properties. 
And this construction is similar to the considerations 
given by Boltzmann in answer to the Zermelo objection:  
if one state is arbitrarily chosen from the set 
of all possible initial states,
then it is most incredible that this will be 
a state with decreasing entropy.  
This notion, which is close to that of stability, 
does not belong to the mathematical theory of probabilities,  
but it must belong to the new revised ``theory of
probabilities'' because it corresponds to the class of problems
mentioned above.  

It should be noted that the common mathematical notions such as
``necessary conditions'' and ``sufficient conditions'' 
are not sufficient for solving the specific problems arising 
in NPPs after accidents.
For example, 
it is possible to write necessary conditions 
for constructing the sarcophagus 
so that a release will necessarily occur, 
but it is more than difficult to write  
``sufficient conditions'' for ensuring 
that no release will occur. 

How can we choose the most reliable ``stable'' construction 
from a set of equivalent constructions?
This problem is close to the Peierls ``law'' 
and to the notion of ``general position.''

The general thermodynamical approach 
with the relationship between the service time and 
the time of relaxation to the quasi-equilibrium state,
(i.e., 
the asymptotics in general rules of averaging) 
taken into account
shows 
that there is something common between such phenomena
as the spouting of superfluid helium 
in Allen--Jones experiments in 1938 
and the radioactive releases in NPPs. 
This is the zeroth-order phase transition, 
which has not been noticed by theoretical physicists \cite{04,05}.
In this phase transition, 
not only the entropy (the intrinsic energy),
but also the thermodynamical Gibbs potential,
has a jump. The instant at which this can occur
can be determined from the corresponding equations of state. 
This effect essentially differs from the well-known picture 
of the first-order phase transition (cf. Fig.~4 in \cite{05}).

A hydrodynamical rather slowly process was also be discovered  
and studied in the rough approximation in~\cite{01}.

It was a common opinion that fuel rods will necessarily melt 
everything under them and go through the ground 
into the subsoil waters as an iron goes through ice.
This could cause an ecological disaster.
To avoid this, even before our calculations were finished, 
a massive concrete basement was built under the debris
in order to keep the fuel rods from going under the ground.
But this enormous work was, in fact, done for nothing,
since the calculations showed that this undesirable effect  
could be avoided. 

The author and the late V.~P.~Myasnikov found that 
this problem can be solved by using the Darcy equation, 
i.e., the equation of filtration. 
To solve this problem, we performed several experiments and 
numerous mathematical calculations. 
It turned out that, even in the case of filtration,  
because of the difference in pressure, 
a small blow from beneath decreases the temperature 
under the fuel rods. This phenomenon is similar to the chimney
effect when the chimney grate is not heated, 
although the temperature of the burning coal is very high.
This was confirmed in practice, none of the fuel rods came 
to the concrete basement.

The three-dimensional calculations taking the construction of
the sarcophagus and the hydrodynamical process of ``convection''
in the filtration medium into account 
were performed by the author more precisely 
in \cite{06,07,08} and in \cite{09}, pp.~245--325.

\section{General concepts concerning\\ 
the zeroth-order phase transition} 

\subsection{Main large parameters in thermodynamics}
 Thermodynamics studies 
steady-state processes in which, 
{\it independently\/} of its initial state,
the system comes to a state that 
remains the same in what follows. 
 But if such a state still varies, 
then this is a thermodynamic variation
only if the process occurs {\it extremely\/}
slowly~\cite{a1}. 
 In other words, 
we slightly vary the state of the system   
and wait until it returns to equilibrium 
and again is independent of the initial state.

 Thus, a large parameter, the time, is invisibly present 
in this case
i.e., 
we observe the system at huge time intervals.
 On the other hand, 
thermodynamics is the limit macroscopic theory
obtained from the microscopic statistical physics
as the number of particles $N$ tends to infinity. 
 Hence there are two large parameters, 
and many things depend on their ratio.
 In turn, the classical statistical physics 
is the limit of quantum statistics as $h\to 0$ 
($h$ is the Planck parameter). 
 Thus, already three large parameters ``meet" here:
time scale, number of particles, and~$1/h$.

In thermodynamics, 
sufficiently large time periods are also often considered,
and in these periods only a part of the system
comes to equilibrium;
this is the so-called ``local equilibrium."  
For example, when plasma is heated by a magnetic field,
the ions begin to obey the Maxwell distribution
after a large observation period, 
and 
the entire system comes to the thermodynamic equilibrium
only after a significantly large time period.

\subsection{The Gibbs postulate}
 Let us consider the Gibbs postulate, 
which has the following form in quantum statistics.

 Suppose that a system is characterized by 
the Hamiltonian operator~$\wh H$ in the Hilbert space 
(in particular, in the Fock space)~$\Phi$. 
 Suppose that the operator~$\wh H$ 
has a nonnegative discrete spectrum
$\lambda_0,\lambda_1,\dots,\lambda_n\dots$\,.
 Then the free energy depending on 
the temperature~$\theta$ is determined as 
$$
E=\theta\ln\sum_{n=0}^\infty e^{-\lambda_n/\theta}\delta_n,
$$
where~$\delta_n$ is the multiplicity 
of the eigenvalue~$\lambda_n$.

 We note that one more parameter appears in the Gibbs postulate,
namely,
$$
E=\theta\ln\lim_{M\to\infty}
\sum_{n=0}^Me^{-\lambda_n/\theta}\delta_n.
$$
 This is one of the most important facts, 
since $\lim_{M\to\infty}$ and~$\lim_{N\to\infty}$ 
do not commute!
 It turns out that 
one must first pass to the limit as $M\to \infty$
and then as $N\to \infty$.

 If we speak about local equilibrium, 
then, as a rule, the Gibbs formula
deals with a subset $\{\lambda_{n'}\}$ of the set $\{\lambda_n\}$.
 This must be done in the study  
of the above problem concerning 
the ions distribution in plasma on heating. 
 This is also done in glass and 
in several other physical problems. 

\subsection{``Friction'' operator}
 At temperature $\theta= 0$,
the Gibbs formula gives 
the value of the lowest eigenvalue, 
which, in physics, is called the ``ground state.''

 This is a law which has not been stated accurately
and which was called the ``energetic efficiency"
by N.~N.~Bogolyubov and others.
 The lower energy is energetically more efficient.
 For example, if a system is perturbed
by an operator $\wh V\:(\Phi\to \Phi)$
and this operator is sufficiently small,
then the transition matrix 
from the state~$\lambda_n$ to the state~$\lambda_m$ 
is determined by the matrix element
$(\Psi_n\wh V\Psi_m^*)$, 
where~$\Psi_n$ and~$\Psi_m$
are the eigenfunctions corresponding 
to~$\lambda_n$ and~$\lambda_m$, 
respectively (see~\cite{a2}).

 The square of this matrix element 
is the probability of transition 
from the state with energy~$\lambda_n$ 
to the state with energy~$\lambda_m$.
 But if $\lambda_n< \lambda_m$, 
then this transition is ``not energetically efficient" 
and hence, 
as if from the viewpoint of statistical physics 
and thermodynamics,  is unrealizable, 
i.e., from the mathematical viewpoint, 
it must be set to be zero.
 Then the transition matrix is not a self-adjoint matrix
with zero entries above the diagonal.
 This means that the friction is taken into account. 
 A pendulum oscillates and stops in the end
if the friction is taken into account, 
i.e., comes to the ``ground state."
  The ``friction" operator was considered 
in more detail in author's paper~\cite{a3}.

\subsection{Phase transitions}
 The derivative of $\epsilon(\theta)$
with respect to~$\theta$ is called the entropy, 
and when the entropy has a jump at a point~$\theta_0$,
it is said that a first-order phase transition occurs; 
when the second-order derivative has a jump,
it is said that a second-order phase transition occurs, 
etc.

 But, of course, not the function itself has a jump, 
but the leading term of its asymptotics as
$N\to \infty$.
 In experiments, 
this ``jump" can sometimes be actually extended in time, 
but we have agreed to consider variations
in sufficiently large periods of time. 
 In these periods,  
not only thermodynamic, but also dynamical processes
can occur. 
 We neglect them and consider only the time periods 
in which the system comes to equilibrium 
with the ``energetic efficiency" taken into account~\cite{a4}.

 The author discovered the zeroth-order phase transition 
both in the theory of superfluidity and superconductivity  
and in economics 
(a stock price break-down, a default, etc.), 
and, quite unexpectedly, it turned out that, 
in view of the natural axiomatics (see~\cite{a5}),
the zeroth-order phase transition 
is related to quantum statistics and thermodynamics.
 This phase transition
has not been noticed by physicists, 
and, perhaps, it contradicts their ideas
that the free energy can be determined 
up to a constant.

\subsection{Metastable state}
 We consider a simple example of 
semiclassical approximation of 
the one-dimensional Schr\"odinger equation
$$
-h^2y_n^{''}+u(x)y_n=\lambda_ny_n, \qquad
y_n(x)\in L_2, \quad h\ll1,
$$
where $u(x)=(x^2-1)^2+qx$ and $q> 0$ is a constant.

 In classical mechanics, 
this equation gives the picture of 
``two cups with a barrier between them.'' 
 If a particle is at the bottom 
of the right-hand cup with higher walls,
then it can get into the other cup with lower walls 
only if the barrier disappears.

 From the viewpoint of the Gibbs postulate,
as $\theta\to 0$,  
the particle must be in the deeper well. 
 Nevertheless, 
it is obvious that if $h\ll 1$, 
then the particle will stay in the well with higher walls
for a very long time. 
 In this case, 
the summation in the Gibbs formula  
must be performed  
over the subset of eigenvalues
corresponding to the well with higher walls 
whose eigenfunctions tend to zero as $h\to0$
outside this well. 
 Moreover, 
the temperature~$\theta$ must be not too high,
so that the eigenvalues above the barrier 
do not play any role in the Gibbs formula.

 Such a state at a local minimum 
of the potential well $u(x)$ 
is an example of a metastable state.

 If we consider the matrix element 
of transition from the lower level
corresponding to the well with higher walls
to the lowest level~$\lambda_0$ 
(corresponding to the bottom of the deep well), 
then it turns out to be exponentially small with respect to~$h$, 
but any transition to higher levels 
is forbidden according to the ``energetic efficiency" law
(i.e., we consider perturbations by the ``friction").

\subsection{Superfluidity}
 N.~N.~Bogolyubov used Landau's ideas
(see the footnote on p.~219 in~\cite{a6})
to show that superfluidity is not a fluid motion,
not a dynamics, 
but a state of fluid such as, for example, 
ice or vapor for water.

 This state corresponds to a metastable state of the system
such that 
any transition from this state to the normal state 
is almost forbidden:  
it is exponentially small as $N\to \infty$.

 N.~N.~Bogolyubov proved this rigorously
under the assumption that 
the system of Schr\"odinger equations is periodic; 
in other words, 
the Schr\"odinger equations were considered on a torus. 
 The spectrum of superfluid velocities 
(the energy levels corresponding to the relevant momenta) 
was discrete.
 This readily distinguishes the state of superfluidity 
from the hydrodynamics of fluids. 
 In the limit as the torus radius tends to infinity, 
the spectrum does not become, as usual, a continuous spectrum, 
but becomes an everywhere dense point spectrum.

 At his time, 
the author also made such a mistake 
studying a problem for the one-dimensional 
Schr\"odinger equation.
 He assumed that the spectrum was continuous, 
while the spectrum turned out to be 
an everywhere dense point spectrum. 
 This concerns the counterexample 
to the Kos\-tyu\-chen\-ko--Morin--Gelfand--Shilov theorem 
given by the author in 1960.
 All of them and all the specialists
agreed with this counterexample 
and found mistakes in their own works.
 Only 20 years later, 
Harris found a ``hole" in the author's proof,
which was announced by Barry Simon in his extensive work. 
 After this, 
S.~Molchanov and the author found 
that the example was faulty 
and the spectrum was not continuous, 
as it seemed to be obvious to everybody,
but it was an everywhere dense point spectrum.
 Only in 1987, 
the author and S.~Molchanov, 
using the same idea,
could find a counterexample 
with a purely continuous spectrum~\cite{a7}.

 In the author's opinion, 
the fact that the spectrum is everywhere dense pointwise
in the limit
can be easily explained from the physical viewpoint. 

 Indeed, if the system is in a state 
with a superfluid velocity~$v$, 
then its transition to a state with larger velocity 
is forbidden 
by the ``energy state efficiency" law,
and its transition to a state with any lesser velocity 
is forbidden 
by the fact that any decrease in the velocity 
contradicts the notion of superfluidity. 
 From the mathematical viewpoint, 
this means that the larger the torus radius, 
the less (exponentially less) 
is the transition from one state to another.

\subsection{Zeroth-order phase transition}
{\bf1.7.1.}
 The author explained the spouting effect
discovered in 1938 by J.~Allen and H.~Jones 
when a superfluid was ``flowing'' 
through a capillary of diameter $10^{-4}$\,cm
(in fact, this superfluid was at the superfluid level 
of a metastable state). 

 The author used the two-level model to show that, 
at a point heated (by light)
till the phase transition temperature~$\theta_c$, 
the heat capacity becomes infinite,
the entropy has a jump, 
and the free energy decreases to its lower value,  
i.e., 
to the point entered by the curve issuing 
from the ground state heated to the temperature~$\theta_c$.
 This means that 
a zeroth-order phase transition occurs. 
 In the present paper, 
we show that the same picture also appears 
in N.~N.~Bogolyubov's model
of a weakly nonideal Bose gas~\cite{a8}.

 This phenomenon can be easily explained 
if we assume that superfluidity is not a thermodynamic state,
but the motion of a fluid. 
 At $\theta= \theta_c$,
the fluid becomes viscous and
cannot penetrate through a thin capillary.

 But the point is that this is not any motion but a state,
and then this is a zeroth-order phase transition.

 The following question arises.
 What will happen to superfluidity 
not in a capillary, but in a rather thick pipe?
 Where is the zeroth-order phase transition?   
 At $\theta= \theta_c$, 
the fluid passes from the superfluid state 
to the state of the usual fluid with viscosity 
and begins to flow according to usual laws 
of hydrodynamics.
 After a while, 
the motion stops, 
and the transition from the superfluid state 
to the state of a fluid at rest 
is a thermodynamic transition. 
 The total intermediate flow is the hydrodynamics,
which must be neglected on our time scale.

{\bf1.7.2.}
 Let us consider the example studied in Sec.~1.5 in more detail.
 We slowly vary the constant~$q$.
 We note that, 
in thermodynamics with the field taken into account, 
there are two more thermodynamic quantities: 
the field strength and the charge. 
 Thus a variation in~$q$ is 
a variation in a thermodynamic variable.

 We show that the passage of~$q$ through the zero point 
results in a zeroth-order phase transition. 
 Indeed, the resonance occurs at $q= 0$: 
the eigenfunctions are already not concentrated 
in one of the wells 
and the probability  
(the square of each eigenfunction) 
is identically distributed over both wells.
 The number of eigenvalues is ``doubled''
and the Gibbs distribution has a jump. 
 Precisely here, 
one can see the role of time.
 Some time is required for half the function,
decreasing outside the first well,
to be ``pumped" into the second well.
 But the Gibbs formula does not take this into account. 
 It may happen that we shall wait 
for this transition for a long time, 
as it was in the preceding example with a thick pipe. 

 If~$q$ becomes negative, 
then all the eigenfunctions remain  
both in the first and in the second well, 
and the Gibbs formula is taken over all the eigenvalues,
rather than over some of them.

{\bf1.7.3.}
 Finally, we consider the effect of transition 
into the turbulent flow for fluid helium, 
which, in fact, is very close to Landau's idea 
concerning the energy pumping 
between large and small vortices.

As will be shown in another paper, 
just the resonance that occurs  
between vortices of these two types
results 
in a zeroth-order phase transition, 
which sharply changes the thermodynamic parameters 
from the viewpoint of thermodynamics
in which we take into account 
the large scales of time between transitions
and, naturally, the averaging over these times.

{\bf1.7.4.}
 Since only the values of bank notes are important, 
while their numbers do not play any role,
and the interchange of two bank notes of the same denomination 
is not an operation, 
the bank notes obey the Bose statistics. 
 The averaging of gains is a nonlinear operation, 
as well as the addition. 
As was shown above,
the only nonlinear ``arithmetics" (the semiring)
that satisfies the natural axioms 
of averaging for Bose particles (bank notes)
has the form $a\oplus b=\ln(e^a+ e^b)$.

 This lead to a formula of Gibbs type. 
 A variation in $\beta=1/T$ 
can be treated as a variation 
in the rouble purchasing power
caused, for example,
by printing a lot of new bank notes. 
 After this, for a period of time, 
the balance (equilibrium) is again established.

 The usual financial efficiency
plays the role of energetic efficiency.
 The zeroth-order phase transition
is either a default or a crisis~\cite{a5,a9}.

\section{An exactly solvable model}
 First, we consider 
the one-dimensional Schr\"odinger equation
for a single particle on the circle
\begin{equation}
\wh H\psi_k(x)=E\psi_k(x), \qquad \psi_k(x-L)=\psi_k(x),
\tag{1}
\end{equation}
where $\psi_k(x)$ is the wave function,  
$x$ takes values on the circle,
and~$\wh H$ is a differential operator of the form 
\begin{equation}
\wh H=\epsilon(-i\hbar\partial/\partial x),
\qquad \epsilon(z)\in C^\infty.
\tag{2}
\end{equation}
 Here~$\hbar$ is the Planck constant, 
$\psi_k(x)=\exp(ip_kx)$, 
where $p_k=2\pi\hbar k/L$, 
$k$ is an arbitrary integer, 
and the corresponding eigenvalues are equal to 
\begin{equation}
E_k=\epsilon(p_k).
\tag{3}
\end{equation}
 We pass from the Hamiltonian function $\epsilon(p)$ 
to a discrete function $\wt\epsilon(p)$ of the form 
\begin{equation}
\wt\epsilon(p)=\epsilon(n\Delta p)
\qquad \text{for}\quad
\Delta p\Bigl(n-\frac12\Bigr)\le p<\Delta p\Bigl(n+\frac12\Bigr),
\tag{4}
\end{equation}
where~$n$ is an arbitrary integer
and~$\Delta p$ is a positive constant.
 Then the Hamiltonian~\thetag{2} changes appropriately,
and we denote the new Hamiltonian by~$\wh{\wt H}$. 
 The eigenfunctions of this operator coincide 
with~$\psi_k(x)$, 
and the eigenvalues~\thetag{3} become
\begin{equation}
\wt E_k=\wt\epsilon(p_k).
\tag{5}
\end{equation}
 In what follows, we assume that the constant~$\Delta p$ 
takes the form 
\begin{equation}
\Delta p=\frac{2\pi\hbar G}L,
\tag{6}
\end{equation}
where~$G$ is a positive integer. 
 In this case, it follows from~\thetag{4}
that the set of energy levels~\thetag{5}
is the set of $G$-fold degenerate energy levels 
\begin{equation}
\lambda_n=\wt E_{Gn}=\epsilon(p_{Gn}).
\tag{7}
\end{equation}

 The Schr\"odinger equation 
for~$N$ noninteracting particles has the form 
\begin{equation}
\wh H_N\Psi(x_1,\dots,x_N)=E\Psi(x_1,\dots,x_N),
\tag{8}
\end{equation}
where $\Psi(x_1,\dots,x_N)$ is a symmetric function
of the variables $x_1,\dots,x_N$ (bosons).
 The Hamiltonian~$\wh H_N$ is given by the formula
\begin{equation}
\wh H_N=\sum_{j=1}^N\wh{\wt H}_j,
\tag{9}
\end{equation}
where $\wh{\wt H}_j$ is the Hamiltonian 
of the particle with number~$j$,
which has the form 
\begin{equation}
\wh{\wt H}_j
=\wt\epsilon\biggl(-i\hbar\frac\partial{\partial x_j}\biggr).
\tag{10}
\end{equation}
 The complete orthonormal system of symmetric eigenfunctions
of the Hamiltonian~\thetag{9} has the form 
\begin{equation}
\Psi_{\{N\}}(x_1,\dots,x_N)
=\Symm_{x_1,\dots,x_N}\psi_{\{N\}}(x_1,\dots,x_N),
\tag{11}
\end{equation}
where~$\Symm_{x_1,\dots,x_N}$ is the symmetrization operator
over the variables $x_1,\dots,x_N$, \ 
$\{N\}$ is the set of nonnegative integers~$N_k$, 
$k\in \ZZ$, satisfying the condition
\begin{equation}
\sum_{k=-\infty}^\infty N_k=N,
\tag{12}
\end{equation}
and the function $\psi_{\{N\}}(x_1,\dots,x_N)$ 
is equal to 
\begin{equation}
\psi_{\{N\}}(x_1,\dots,x_N)
=\prod_{s=1}^N\psi_{k_s}(x_s).
\tag{13}
\end{equation}
 Here the indices $k_1,\dots,k_N$
are expressed in terms of the set $\{N\}$ 
using the conditions
\begin{equation}
k_s\le k_{s+1} \quad \text{for all}\ 1\le s\le N-1,
\qquad
\sum_{s=1}^N\delta_{kk_s}=N_k \quad \text{for all}\ k\in\ZZ,
\tag{14}
\end{equation}
and~$\delta_{kk'}$ is the Kronecker symbol. 
 The eigenvalues of the Hamiltonian~\thetag{9} are
\begin{equation}
E(\{N\})=\sum_{k=-\infty}^\infty\wt E_kN_k.
\tag{15}
\end{equation}
 We consider the interparticle interaction 
of the following form. 
 We assume that the particles interact in pairs 
and the interaction operator for particles with numbers~$j$
and~$k$ has the form 
\begin{equation}
\wh V_{jk}=-\frac VNW\bigl(\wh{\wt H}_j-\wh{\wt H}_k\bigr),
\tag{16}
\end{equation}
where $V> 0$ is the interaction parameter 
and the function $W(\xi)$ is given by the formula
\begin{equation}
W(\xi)=\begin{cases}
1& \text{for}\ |\xi|<D,\\
0& \text{for}\ |\xi|\ge D.
\end{cases}
\tag{17}
\end{equation}
 Here $D> 0$ is the parameter of the width of interaction
with respect to energy.
 The operator~\thetag{17} corresponds to the interaction
under which particles in a pair attract each other
and radiate the energy quantum $-V/N$ 
if the difference between their energies is less than~$D$
and do not interact at all 
if the difference between their energies is larger than~$D$.
 The Hamiltonian of the system of~$N$ particles 
with interaction~\thetag{17} has the form 
\begin{equation}
\wh H_N=\sum_{j=1}^N\wh{\wt H}_j
+\sum_{j=1}^N\sum_{k=j+1}^N\wh V_{jk}.
\tag{18}
\end{equation}
 In view of~\thetag{16}, the sums in~\thetag{18} commute, 
and hence the set of eigenfunctions 
of the Hamiltonian~\thetag{18} coincides with~\thetag{11}.
 It also follows from~\thetag{16} 
that the corresponding eigenvalues are equal to 
\begin{equation}
\EE(\{N\})
=\sum_{k=-\infty}^\infty\wt E_kN_k
-\frac V{2N}\sum_{k=-\infty}^\infty
\sum_{l=-\infty}^\infty
W(\wt E_k-\wt E_l)(N_kN_l-\delta_{kl}N_k).
\tag{19}
\end{equation}
 In what follows, we assume that 
the interaction width is sufficiently small 
and satisfies the condition
\begin{equation}
D<\min_{n\ne m}|\lambda_n-\lambda_m|.
\tag{20}
\end{equation}
 The set of energy levels~$\wt E_k$, $k\in \ZZ$,
coincides with the $G$-fold degenerate set of
levels~\thetag{7}.
 Hence, by~\thetag{20}, 
the set of energy levels~\thetag{19} 
of the system of~$N$ particles under study
can be written as 
\begin{equation}
\EE(\{\wt N\})
=\sum_{n=-\infty}^\infty\lambda_n\wt N_n
-\frac V{2N}\sum_{n=-\infty}^\infty\wt N_n(\wt N_n-1),
\tag{21}
\end{equation}
where the level $\EE\{\wt N\})$ has the multiplicity 
\begin{equation}
\Gamma(\{\wt N\})
=\prod_{n=-\infty}^\infty
\frac{(G+\wt N_n-1)!}{(G-1)!\,\wt N_n!}.
\tag{22}
\end{equation}
 Here $\{\wt N\}$ denotes the set of nonnegative 
integers~$\wt N_n$, $n\in \ZZ$, satisfying the condition
\begin{equation}
\sum_{n=-\infty}^\infty\wt N_n=N.
\tag{23}
\end{equation}

 Let us consider the statistical sum for 
the system of~$N$ bosons with Hamiltonian~\thetag{18}.
 Since the energy levels and their multiplicities
are given by formulas~\thetag{21} and~\thetag{22},
respectively,
the statistical sum at temperature~$\theta$ 
takes the form  
\begin{equation}
Z(\theta,N)
=\sum_{\{\wt N\}}\Gamma(\{\wt N\})
\exp\bigl((-\EE\{\wt N\})\theta\bigr).
\tag{24}
\end{equation}
 Here the summation is performed over all sets  
$\{\wt N\}$ with condition~\thetag{23} taken into account.

 In what follows, we assume that~$G$ depends on~$N$
and the following condition is satisfied:
\begin{equation}
\lim_{N\to\infty}\frac GN=g>0.
\tag{25}
\end{equation}

 By $\wt F(\{\wt N\},\theta)$ 
we denote a function of the form 
\begin{equation}
\wt F(\{\wt N\},\theta)
=\EE(\{\wt N\})-\theta\ln\bigl(\Gamma(\{\wt N\})\bigr),
\tag{26}
\end{equation}
and by $\{\wt N^0\}$ the set of nonnegative numbers~$\wt N_n^0$,
$n\in \ZZ$, 
for which the function~\thetag{26} 
is minimal under condition~\thetag{23}.

 Now we consider the problem of finding the minimal value
of the function~\thetag{26} under condition~\thetag{23}.
 In the limit as $N\to \infty$ and under condition~\thetag{25}, 
the point of minimum has the form 
\begin{equation}
\wt N_n(\theta)=N\bigl(m_n(\theta)+o(1)\bigr),
\tag{27}
\end{equation}
where $m_n(\theta)$, $n\in \ZZ$, 
is determined by the system of equations
\begin{equation}
\lambda_n-Vm_n+\theta\ln\biggl(\frac{m_n}{g+m_n}\biggr)
=\mu(\theta), \qquad n\in\ZZ,
\tag{28}
\end{equation}
and $\mu(\theta)$ can be found from the equation
\begin{equation}
\sum_{n=-\infty}^\infty m_n=1.
\tag{29}
\end{equation}
 The substitution of~\thetag{27} into~\thetag{26} 
and then the use of the asymptotic Stirling formula 
give the following relation for the specific free energy:
\begin{align}
f(\theta)
&\equiv\lim_{N\to\infty}f(\theta,N)
=\lim_{N\to\infty}\frac{\wt F(\{\wt N^0\},\theta)}N
\notag\\ 
&=\sum_{n=-\infty}^\infty\biggl(\lambda_nm_n-\frac V2m_n^2\biggr)
+\biggl(\theta m_n\ln\biggl(\frac{m_n}g\biggr)
-\theta(g+m_n)\ln\biggl(1+\frac{m_n}g\biggr)\biggr),
\tag{30}
\end{align}
where, for brevity, 
we omit the argument~$\theta$ of~$m_n(\theta)$.
 We introduce the notation 
\begin{equation}
\omega_n=\lambda_n-Vm_n.
\tag{31}
\end{equation}

 In the notation~\thetag{31}, 
the system of Eqs.~\thetag{28} and~\thetag{29} 
takes the form 
\begin{gather}
\omega_n(\theta)
=\lambda_n-V\frac g{\exp((\omega_n-\mu)/\theta)-1},
\qquad n\in\ZZ,
\tag{32}
\\
\sum_{n=-\infty}^\infty\frac g{\exp((\omega_n-\mu)/\theta)-1}=1.
\tag{33}
\end{gather}
 The system of Eqs.~\thetag{32} and~\thetag{33} 
{\it exactly\/} coincides 
with the temperature Hartree equations~\cite{a10}
for the system of~$N$ bosons with Hamiltonian~\thetag{18}.

 We shall study the solutions 
of Eqs.~\thetag{28} and~\thetag{29}.
 For $\theta= 0$, 
the system has many solutions 
which we number by the integer~$l$:
\begin{equation}
m_n^{(l)}=\delta_{ln}, \qquad n,l\in\ZZ.
\tag{34}
\end{equation}
 Among all the numbers~$l$,
we choose those that satisfy the condition
\begin{equation}
\nu_{nl}\equiv\lambda_n-\lambda_l+V>0
\qquad \text{for all}\quad n\ne l.
\tag{35}
\end{equation}
 For these numbers, there exist solutions 
of Eqs.~\thetag{28} and~\thetag{29}
converging to~\thetag{34} as $\theta\to 0$.
 The asymptotics of these solutions as $\theta\to 0$ 
has the form 
\begin{equation}
m_n^{(l)}\sim g\exp\biggl(-\frac{\nu_{nl}}\theta\biggr)
\quad \forall n\ne l,
\qquad
1-m_l^{(l)}\sim g\sum_{n\ne l}
\exp\biggl(-\frac{\nu_{nl}}\theta\biggr).
\tag{36}
\end{equation}
 Thus, depending on the spectrum~$\lambda_n$, $n\in \ZZ$, 
for sufficiently small values of the temperature~$\theta$,
the system of Eqs.~\thetag{28} and~\thetag{29} 
has many solutions. 
These solutions, 
along with the point of global minimum, 
also contain points of local minimum 
of the function~\thetag{26}.
 The values of the function~\thetag{26} 
at the points of local minimum are equal to 
the free energy of metastable states. 
 We consider the function~\thetag{26} at $\theta= 0$.
 In this case, it coincides with the energy spectrum 
of system~\thetag{21}.
 We consider the energy of the system 
for the case in which almost all particles 
are at the energy level~$\lambda_l$, 
which means that following conditions holds:
\begin{equation}
\wt N_n\ll N \qquad \forall n\ne l.
\tag{37}
\end{equation}
 Deriving~$\wt N_l$ from relation~\thetag{23} 
and substituting the result into~\thetag{21}, 
we see that, in view of~\thetag{37}, 
the energy spectrum of the system in this domain 
has the form 
\begin{equation}
\EE(\wt N)
\approx\lambda_lN-\frac{VN}2
+\sum_{n\ne l}(\lambda_n-\lambda_l+V)\wt N_n.
\tag{38}
\end{equation}
 
 To the Hamiltonian~\thetag{18}, there correspond the Hartree
equation and the system of variational equations. 
 To each $l\in \ZZ$, there corresponds 
a solution of the Hartree equation,
and this solution describes the state
\begin{equation}
\wt N_n^{(l)}=N\delta_{nl}.
\tag{39}
\end{equation}
 Moreover, 
the eigenvalues of the system of variational equations
corresponding to this solution of the Hartree equation
coincide with 
\begin{equation}
\nu_{nl}=\lambda_n-\lambda_l+V, \qquad n\ne l.
\tag{40}
\end{equation}
 In~\cite{a11,a12},
it was shown that 
if the eigenvalues of the system of variational equations 
for the solution of the Hartree equation 
are real and nonnegative, 
then such a solution corresponds to the ground state
or a metastable state of the system. 
 This means that to all~$l$ 
for which condition~\thetag{35} holds at $\theta= 0$,
there correspond metastable states of the system of bosons. 
 As in the case of zero temperature, 
to solutions of Eqs.~\thetag{28} and~\thetag{29} 
for $\theta\ne 0$,
there corresponds a temperature metastable state 
if the point~\thetag{27} is a point of local minimum. 
 Now we note that, 
for very large temperatures, 
the system of Eqs.~\thetag{28} and~\thetag{29} 
has only one solution corresponding 
to the global minimum of the function~\thetag{26}.
 The asymptotics of this solution as
$\theta\to \infty$ has the form 
\begin{equation}
n_m(\theta)
\sim g\frac{e^{-\lambda_m/\theta}}
{\sum_{l=-\infty}^\infty e^{-\lambda_l/\theta}}.
\tag{41}
\end{equation}
 The uniqueness of the solution at large temperatures 
means that all metastable states disappear 
with increasing temperature. 
 The temperature at which a metastable state disappears 
is critical for this state.

 Let us consider the behavior 
of the entropy and the heat capacity 
of metastable states 
when we approach the critical temperature. 
 We consider the metastable state
to which there corresponds a solution 
of Eqs.~\thetag{28} and~\thetag{29} 
converging to~\thetag{34} as $\theta\to 0$ 
for some~$l$ for which~\thetag{35} holds.
 In what follows, 
we assume that the function $\epsilon(p)$ 
satisfies the condition 
$\epsilon(0)<\epsilon(p)$ for $p\ne 0$.
 Then, according to~\thetag{7}, 
we have $\lambda_0< \lambda_l$ for $l\ne 0$.
 Hence the solution of 
Eqs.~\thetag{28} and~\thetag{29} converging to~\thetag{34}
as $\theta\to 0$ for $l= 0$
corresponds to the temperature ground state 
of the system of~$N$ bosons. 
 Moreover, 
condition~\thetag{35} becomes equivalent to 
the condition $\lambda_l-\lambda_0< V$.

 We assume that this inequality holds for $l\ne 0$. 
 The condition 
that the corresponding solution 
$m_n^{(l)}(\theta)$ of Eqs.~\thetag{28} and~\thetag{29} 
determines 
the point of local minimum of the function~\thetag{26}
under condition~\thetag{23} by formula~\thetag{27} 
can be written as the system of inequalities
\begin{gather}
\alpha_n^{(l)}(\theta)
\equiv-V+\frac{\theta g}
{m_n^{(l)}(\theta)\bigl(g+m_n^{(l)}(\theta)\bigr)}>0
\quad \forall n\ne l,
\tag{42}
\\
\alpha_l^{(l)}(\theta)
\equiv-V+\frac{\theta g}
{m_l^{(l)}(\theta)\bigl(g+m_l^{(l)}(\theta)\bigr)}<0,
\qquad
-\sum_{n\ne l}\frac{\alpha_l^{(l)}(\theta)}
{\alpha_n^{(l)}(\theta)}<1.\notag
\end{gather}
 We note that these inequalities hold for~\thetag{36}
as $\theta\to 0$.
 Inequalities~\thetag{42} follow from the condition
that the second variation of the function~\thetag{26}
is positive;
the variation is calculated under condition~\thetag{23}.

 The metastable state disappears at the temperature
at which the last inequality in~\thetag{42} 
becomes an equality. 
 We denote this critical temperature by~$\theta^{(l)}_c$.
 It follows from~\thetag{42} and Eqs.~\thetag{28}
and~\thetag{29} for $\theta< \theta_c^{(l)}$
that $m_n^{(l)}(\theta)$ is an increasing function 
of the variable~$\theta$ for $n\ne l$
and
$m_l^{(l)}(\theta)$ is a decreasing function of~$\theta$.
 Moreover, we see that 
$m_l^{(l)}(\theta)>m_n^{(l)}(\theta)>m_{n'}^{(l)}(\theta)$
if $\lambda_n< \lambda_{n'}$ and $n,n'\ne l$.

 From~\thetag{30} we obtain the following expression 
for the specific entropy of a metastable state 
in the limit as $N\to \infty$:
\begin{equation}
s^{(l)}(\theta)
=\sum_{n=-\infty}^\infty\biggl(\bigl(g+m_n^{(l)}\bigr)
\ln\biggl(1+\frac{m_n^{(l)}}g\biggr)
-m_n^{(l)}\ln\biggl(\frac{m_n^{(l)}}g\biggr)\biggr),
\tag{43}
\end{equation}
where, for brevity, we omit the argument~$\theta$
of $m_n^{(l)}(\theta)$.
 Differentiating~\thetag{43}, we obtain
\begin{equation}
\frac{\partial s}{\partial\theta}
=\sum_{n\ne l}\frac{\partial m_n^{(l)}}{\partial\theta}
\ln\biggl(\frac{g+m_n^{(l)}}{m_n^{(l)}}
\,\frac{m_l^{(l)}}{g+m_l^{(l)}}\biggr)
>0.
\tag{44}
\end{equation}
 The last inequality follows from the properties
of $m_n^{(l)}(\theta)$.
 Since
the last inequality in~\thetag{42} becomes an equality
at the critical temperature, 
we can show that, as $\theta\to\theta_c^{(l)}- 0$,
the solutions 
of Eqs.~\thetag{28} and~\thetag{29} behave as follows:
\begin{equation}
m_n^{(l)}(\theta)-m_n^{(l)}(\theta_c^{(l)})
\approx\frac{C^{(l)}}{\alpha_n^{(l)}(\theta_c^{(l)})}
\sqrt{\theta_c^{(l)}-\theta}
\qquad \forall n\in\ZZ,
\tag{45}
\end{equation}
where~$C^{(l)}$ is a negative number.
 The substitution of~\thetag{45} into~\thetag{43}
shows that the derivative of the specific entropy 
with respect to the temperature 
(this quantity is equal to the heat capacity
divided by the temperature)
tends to infinity as
$\theta\to\theta_c^{(l)}- 0$ 
according to the law
$1/\sqrt{\theta_c^{(l)}-\theta}$. 
 This means that the projection 
of a Lagrangian manifold, 
corresponding to a metastable state, 
on the $\theta$-axis is not uniquely determined
in a neighborhood of the critical temperature~\cite{a4,a13}.
 The derivative of the temperature with respect to the entropy
vanishes as the critical temperature is approached.
 Therefore, as well as in view of~\thetag{44},
as was already pointed out,
the Lagrangian manifold is uniquely projected 
on the $s$-axis.
 We note that it follows from the properties of 
$m_n^{(l)}(\theta)$ that 
the following conditions hold for
$\theta< \theta_c^{(l)}$:
\begin{align}
m_0^{(l)}(\theta)<m_n^{(l)}(\theta)
&\qquad\forall n\ne0,l,
\tag{46}
\\
\alpha_0^{(l)}(\theta)<\alpha_n^{(l)}(\theta)
&\qquad\forall n\ne0,l.\notag
\end{align}
 Inequalities~\thetag{46} mean that the potential barrier
between the energy levels~$\lambda_l$ and~$\lambda_0$
is less than 
the energy barrier between~$\lambda_l$ and~$\lambda_n$
for $n\ne0,l$.
 This means that,
as the critical temperature is attained,  
the system of bosons under study 
which is in the metastable state with number~$l$
changes its state in a jump 
and passes to the temperature ground state.
This is a zeroth-order phase transition, 
since not only the entropy and heat capacity, 
but also the free energy have jumps.

 If the quantity  
$\min_{n\ne0}|\lambda_n-\lambda_0|= \delta$ is small,
then the difference between the free energies 
in the zeroth-order phase transition 
from the lowest metastable state to the ground state
is also small, 
and the heat capacity in this transition 
has a singularity. 
 The asymptotics of the statistical sum 
near the critical point 
is given by the 
asymptotics of the canonical operator 
in a neighborhood of a focal point.
 This asymptotics has the form of an Airy type function
of an imaginary argument  
and has a singularity as $N\to \infty$.
 The parameters $\delta\ll 1$, $N\gg 1$, and $L\gg 1$ 
can be chosen so that the form 
of the phase transition point
coincides exactly with the
$\lambda$-point~\cite{a14}--\cite{a24}.

\section{Critical temperature and the phase transition
for a weakly nonideal Bose gas}
 Following the Bogolyubov paper~\cite{a8},
we consider the system of $N$ bosons 
on a three-dimensional torus with sides of length~$L$.
 Suppose that the bosons interact pairwise 
and $\epsilon V(x- y)$ is the potential of this interaction,
i.e., a function on this torus, and
$\epsilon=1/N$. 
 At the temperature~$\theta$, 
the Bogolyubov variational method gives 
the following expression for the free energy
of such a system%
\footnote{The same equations are obtained from the asymptotics 
of ultra-secondary quantized equations of 
thermodynamics~\cite{a10}.}:
\begin{equation}
F(\theta,k_0)=F(\theta,k_0,\{N(\theta)\}),
\tag{47}
\end{equation}
where $F(\theta,\{N\})$ is the function
\begin{align}
&
F(\theta,k_0,\{N\})
=\frac{\hbar^2k_0^2}{2m}N
+\frac\epsilon{2L^3}\wt V(0)N(N-1)
+\sum_{k\ne
k_0}\biggl(\frac{\hbar^2k^2}{2m}-\frac{\hbar^2k_0^2}{2m}\biggr)N_k
\tag{48}\\ 
&\qquad
+\frac\epsilon{L^3}\sum_{k\ne k_0}\bigl(\wt V(0)+\wt V(k-k_0)\bigr)
\biggl(NN_k-2N_k^2-\sum_{k'\ne k_0}N_kN_{k'}\biggr)\notag\\ 
&\qquad
+\frac\epsilon{2L^3}\sum_{k\ne k'\ne k_0}
\bigl(\wt V(0)+\wt V(k-k')\bigr)N_kN_{k'}\notag\\ 
&\qquad
+\frac\epsilon{L^3}\sum_{k\ne k_0}\wt V(0)N_k^2
+\frac\epsilon{2L^3}\wt V(0)\biggl((1-2N)\sum_{k\ne k_0}N_k\notag\\
&\qquad\qquad
+\sum_{k\ne k'\ne k_0}N_kN_{k'}+2\sum_{k\ne k_0}N_k^2\biggr)\notag\\ 
&\qquad
+\theta\sum_{k\ne k_0}\bigl(N_k\ln(N_k)-(N_k+1)\ln(N_k+1)\bigr).
\notag
\end{align}
 Here~$m$ is the mass of bosons, 
$\hbar$ is the Planck constant, 
$k$ and~$k'$ are arbitrary three-dimensional vectors of the form 
$2\pi/L(n_1,n_2,n_3)$, \
$n_1,n_2,n_3$ are integers,
$k_0$ is a distinguished vector of this form, 
and~$\wt V$ is given by the formula
\begin{equation}
\wt V(k)=\int dx\,V(x)e^{-ikx}.
\tag{49}
\end{equation}
 Finally, $\{N\}$ denotes a set of variables~$N_k$,
$k\ne k_0$, and $\{N(\theta)\}$ is the minimum point of the
function~\thetag{48}.
 We note that the function~\thetag{48}
corresponds to the case in which 
the system of bosons contains a condensate 
consisting of particles with momentum~$\hbar k_0$.
 Suppose that $k_0\ne 0$ is less than the Landau velocity, 
i.e., $k_0$ must satisfy the condition
\begin{equation}
\min_{k\ne k_0}\biggl(\frac{\hbar^2k^2}{2m}
-\frac{\hbar^2k_0^2}{2m}+\frac{\wt V(k-k_0)}{L^3}\biggr)
>0.
\tag{50}
\end{equation}
 To such a~$k_0$, there corresponds a metastable state.
 The equations for determining the minimum point 
of the function~\thetag{48},
\begin{equation}
\frac{\partial F(\theta,\{N\})}{\partial N_k}=0
\qquad \forall k\ne k_0,
\tag{51}
\end{equation}
are equivalent to the system of the temperature Hartree equations. 
 In view of~\thetag{50}, Eqs.~\thetag{51} have solutions 
for sufficiently small temperatures~$\theta$,
and the asymptotics as $\theta\to 0$
of these solutions has the form
\begin{equation}
N_k\approx e^{-\omega_{kk_0}/\theta}
\qquad \forall k\ne k_0,
\tag{52}
\end{equation}
where
\begin{equation}
\omega_{kk_0}
=\frac{\hbar^2k^2}{2m}-\frac{\hbar^2k_0^2}{2m}
+\frac{\wt V(k-k_0)}{L^3}.
\tag{53}
\end{equation}
 We note that the substitution of~\thetag{52} 
into~\thetag{48} leads to an expression 
different from the free energy obtained 
by the summation over the quasiparticles.
 Suppose now that~$k_0$ is less than the Landau velocity, 
but still is very close to it.
 This means that the minimum in~\thetag{50}
is attained at $k= 0$, 
and this minimum tends to~$0$ as $N\to\infty$.
 To be definite, we assume that 
\begin{equation}
\omega_{0k_0}=\frac\alpha{N^\sigma},
\tag{54}
\end{equation}
where $1/2< \sigma$.
 Since~$\omega_{kk_0}$ for $k\ne 0$
is a quantity of the order of~$1$, 
it follows from~\thetag{54} that 
\begin{equation}
N_k\ll N_0
\tag{55}
\end{equation}
at any temperature.
 In view of~\thetag{55}, in the limit as $N\to \infty$,
the leading contribution to the function~\thetag{48}
is given by the terms containing~$N_0$;
thus, the terms that do not contain~$N_0$
can be neglected.
 In this case, the system of Eqs.~\thetag{51} 
for determining the minimum point 
becomes the equation 
\begin{equation}
\frac\alpha{N^\sigma}+\frac\beta NN_0-\frac\theta{N_0}\approx0,
\tag{56}
\end{equation}
where
\begin{equation}
\beta=2(2\wt V(k_0)+\wt V(0))>0.
\tag{57}
\end{equation}
 Equation~\thetag{56} can trivially be reduced 
to the quadratic equation.
 An analysis of this equation shows that 
it has two solutions for $0<\theta< \theta_c$, 
where
\begin{equation}
\theta_c=\frac{\alpha^2}{4\beta N^{2\sigma-1}},
\tag{58}
\end{equation}
and Eq.~\thetag{56} does not have solutions 
for $\theta_c< \theta$.

 As $\theta\to\theta_c- 0$, 
the solutions of Eq.~\thetag{56} have the form 
\begin{equation}
N_0(\theta)\approx\sqrt{\frac{N\theta_c}\beta}
\biggl(1-\sqrt{1-\frac\theta{\theta_c}}\biggr).
\tag{59}
\end{equation}
 The substitution of~\thetag{59} into~\thetag{48}
gives the free energy whose second-order derivative 
becomes infinite  
as $\theta\to\theta_c- 0$.
 This means that 
the heat capacity of a metastable state
increases significantly near the temperature 
that is critical for this metastable state:
\begin{equation}
C(\theta,k_0)
=-\theta\frac{\partial^2F(\theta,k_0)}{\partial\theta^2}
\approx\frac1{2\sqrt{\theta_c(\theta_c-\theta)}}.
\tag{60}
\end{equation}
 Thus, it turns out that 
the properties of metastable states in the Bogolyubov model 
are similar 
to the properties of metastable states in the two-level model.
 Namely, 
metastable states disappear at the critical temperature, 
and the heat capacity of these states 
increases according to the law $1/\sqrt{\theta_c-\theta}$
as the critical temperature is approached.

 Hence
we have the same picture of the zeroth-order phase transition 
and the spouting effect
for a weakly nonideal Bose gas in the Bogolybov model.

\end{document}